\documentclass[prb,twocolumn,amsmath,amssymb,citeautoscript]{revtex4}
\usepackage{graphicx}
\usepackage{amsmath}

\def \nn{\nonumber}

\def \half{{\frac{1}{2}}}

\def \yd{{^\dagger}}

\begin{document}
\title{Phase Transitions of $S=1$ Spinor Condensates in an Optical Lattice}
\author{Daniel Podolsky$^{1,2}$, Shailesh Chandrasekharan$^3$
and Ashvin Vishwanath$^{2,4}$} 
\affiliation{ $^1$Department of Physics, Technion, 32000 Haifa, Israel
\\
$^2$Department of
Physics, University of California, Berkeley, CA 94720
\\
$^3$Department of Physics, Box 90305, Duke University, Durham, NC
27708
\\
$^4$Materials Sciences Division, Lawrence Berkeley National
Laboratory, Berkeley, CA 94720 }
\date{Printed \today}

\begin{abstract}
We study the phase diagram of spin-one polar condensates in a two
dimensional optical lattice with magnetic anisotropy. We show that 
the topological binding of vorticity to nematic disclinations allows 
for a rich variety of phase transitions. These include 
Kosterlitz-Thouless-like transitions with a superfluid stiffness 
jump that can be experimentally tuned to take a {\em continuous} 
set of values, and a new {\em cascaded Kosterlitz-Thouless} transition, 
characterized by {\em two} divergent length scales.
For higher integer spin bosons $S$, the thermal phase transition out of
the planar polar phase is strongly affected by the parity of S.
\end{abstract}
\maketitle

\section{Introduction}
\label{sec:intro}

Low temperature superfluidity in simple atomic systems 
is a well-studied subject. It is well-known that scalar bosons, confined to two spatial 
dimensions ($2d$), cannot form a true condensate. Instead, 
the off-diagonal correlations display either power law 
decay (in the superfluid phase) or exponential decay (in the
normal phase). These two phases are usually separated by a
finite temperature Kosterlitz-Thouless (KT) driven by 
superfluid vortex unbinding. Such a transition in atomic 
systems has been observed experimentally through interference 
measurements \cite{Hadzibabic}. 

Atomic systems of bosons with degenerate hyperfine levels lead to more
exotic phenomena at low temperatures usually referred to as 
``spinor condensation''. Such systems have been at the focus 
of intense experimental and theoretical activity since their 
discovery \cite{stenger}. The hyperfine levels give rise to 
a new quantum number analogous to the spin. The macroscopic 
phase coherence in spinor systems can be accompanied by magnetic 
order. Indeed the spin and charge degrees of freedom may be 
strongly intertwined -- as seen, for example, in the topological 
defects, which can simultaneously involve atomic supercurrents 
and magnetic textures \cite{Ho}. The presence of these multiple 
types of defects leads to a richer variety of phases and phase 
transitions.
%What effect does the coupling of spin and
%charge have on the phase transitions in these systems?

In this paper, we study thermal transitions in $2d$ polar 
condensates, where uniaxial spin nematic order (characterized by 
a headless vector, or ``director'' ) coexists with superfluidity. 
First, we point out a crucial difference between polar condensates with 
even integer and odd integer spin (as in $S=1$ $^{23}$Na). For 
even $S$, the superfluid
vortex and the nematic disclination are independent, while for
odd $S$ they are bound to each other topologically. This strongly
impacts the phase diagrams. When the nematic director is confined
to rotate in a plane, the normal state can be reached via a single
continuous transition from the polar state for the case of odd
$S$, but not for even $S$, where a split transition is
expected. We study the odd $S$ case in detail in this paper,
and further show that the single continuous transition itself can
take on two very different characters, one, which is essentially
KT-like, but with a non-universal superfluid stiffness jump; and
another that is of a new `cascaded-KT' type described
in more detail below.   Interference experiments \cite{Hadzibabic} 
which have been used to study KT transitions in scalar condensates, 
can also be used to probe the new transitions discussed here.

%These different transitions are of relevance
%when spin and charge stiffnesses are tuned separately in an optical lattice.
% outside the KT universality class.

% Mukerjee {\em et al.} showed that the
%interaction of spin and charge in isotropic spin-1 polar
%condensates leads to a KT transition with a jump in superfluid
%stiffness that is four times as large as in the conventional case
%\cite{MukerjeeXuMoore}. Here, we look at spin-1 polar condensates

Consider spin-one bosons in a 2d optical lattice, described by a Hubbard model 
with couplings $U_0$ and $U_2$,
\begin{eqnarray}
{\cal H}&=&-t\sum_{\langle ij \rangle,\sigma}a_{i\sigma}\yd
a_{j\sigma}+U_0\sum_i n_i(n_i-1)-\mu\sum_i n_i\nn\\
&\,&+U_2
\sum_i(\vec{S}_i^2-2n_i)-g\sum_i(S_i^z)^2.\label{eq:Hubbard}
\end{eqnarray}
The depth of the optical lattice serves to tune $(U_0/t)$.
Here $a_{i\sigma}^\dagger$ creates an atom at site $i$ with spin
$S^z_i=\sigma\in\{-1,0,1\}$, $n_i$ is the
particle number at site $i$, and $\mu$ is the chemical
potential. The quadratic Zeeman field $g$, described below, is
absent in magnetically isotropic systems.  We concentrate on atoms
with antiferromagnetic spin interactions, $U_2>0$, {\it e.g.} as is 
the case in $^{23}$Na.

The zero temperature phase diagram of model (\ref{eq:Hubbard})
with $U_2>0$ and $g=0$ was studied in Ref.~[\onlinecite{Imambekov}]. At
unit filling (one atom per lattice site), the system undergoes a
continuous transition at $T=0$ between a nematic Mott insulator
and a polar superfluid.  The transition is tuned by the depth of the optical 
lattice. For deep lattices $(U_0/t\gg 1)$ the system is a
nematic Mott insulator, characterized by atoms which predominantly occupy 
the $S_z=0$ state, together with a vanishing compressibility.  More 
generally, the nematic can be any $\hat{n}\cdot S=0$ state,
with the director $\hat{n}$ serving as an order
parameter. 
%Note that the nematic does not break time-reversal
%symmetry, $\langle \vec{S}_i\rangle=0$, but it does break
%rotational symmetry, since $\langle
%(\hat{m}\cdot\vec{S}_i)^2\rangle$ vanishes only when
%$\hat{m}=\hat{n}$.
On the other hand, for weak optical lattices $(U_0/t\ll 1)$ the
system is a polar superfluid. The order parameter
$\Psi_\sigma\equiv\langle a_\sigma
\rangle=(\psi_{+1},\psi_0,\psi_{-1})^T$ in the polar state is ${\bf
\Psi}=e^{i\theta}{\cal R}(0,1,0)^T$ -- ${\cal R}$ is a generic
$SO(3)$ spin rotation. As in the Mott insulator, the polar
state $\bf{\Psi}$ has nematic order, described by a director
${\hat{n}}$ for which ${\hat{n}\cdot\vec{S}}|{\bf\Psi}\rangle=0$,
but it {\em also} has superfluid order, captured by an expectation
value of the superfluid phase $\theta$.

\begin{figure}
\includegraphics[width=2.6 in]{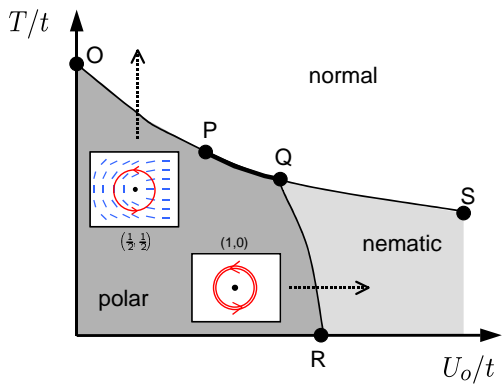}
\caption{Schematic phase diagram as a function of
optical lattice depth and temperature. 
The topological defects that disorder the polar
state are: a superfluid vortex $(q_c,q_s)=(1,0)$, where $\theta$
winds by $2\pi$ (red double circle);  and a
disclination+half-vortex $(\half,\half)$, where both $\theta$ and
$\phi$ wind by $\pi$. Along the cascaded KT transition {\bf PQ}, both defects
play a role.
\label{fig:TvsU} } 
\end{figure}

In the following, we are interested in systems with positive
quadratic Zeeman field $g>0$. Such a field has the effect of
restricting the director $\hat{n}$ to lie in the $xy$-plane in
both the Mott nematic and polar states\cite{stenger}. The most
general planar polar state is then
\begin{eqnarray}
{\bf\Psi}=e^{i\theta}\left(-e^{i\phi},0,e^{-i\phi}\right)^T,\label{eq:planar}
\end{eqnarray}
where $\phi$ is the angle of $\hat{n}$ relative to the
$x$-axis.  The AC Zeeman effect -- shining linearly polarized
light at a frequency slightly detuned from the hyperfine level
splitting -- can induce the required negative quadratic Zeeman
field \cite{DSK} that leads to a planar polar state. The opposite
$g<0$ limit,  (induced by a magnetic field in S=1 $^{23}$Na) is
essentially identical to a non-magnetic system since the nematic
director is frozen along the field.

\section{Topological defects}
\label{sec:defects}

Topological defects play an essential
role in $2d$ finite temperature continuous phase transitions.  In
the present context, the planar polar SF is the ``most ordered''
phase, as it has both nematic and superfluid quasi long range
order (QLRO). We can then understand the phase diagram in terms of
proliferation of defects of the planar polar condensate, which can
destroy the order partially (leading to a nematic insulator) or
completely.

It is impossible to distinguish between the states
$\pm\hat{n}$ ($\hat{n}$ is a headless vector).  However, an adiabatic
rotation taking $\hat{n}$ to $-\hat{n}$ induces a change of sign
in the polar order parameter ${\bf \Psi}$. This sign can be
absorbed by simultaneously shifting the superfluid phase $\theta$
by $\pi$. Therefore, to insure single-valuedness of the order
parameter, {\it in a polar state, a nematic disclination must be
accompanied by a half-vortex in the superfluid phase}.

\begin{figure}
\includegraphics[width=2.8 in]{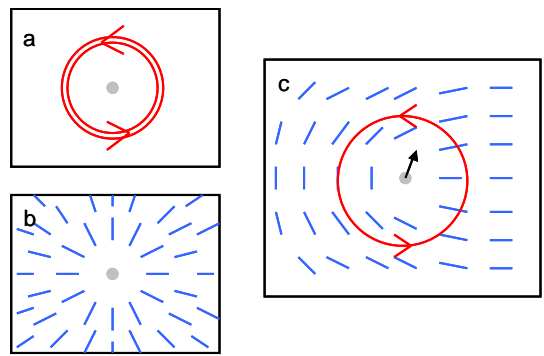}
\caption{Topological defects of the planar polar state. (a) In a
superfluid vortex $(q_c,q_s)=(1,0)$, $\theta_c$ winds by $2\pi$
around the vortex (superflow shown as double circle with arrows).
(b) In a nematic vortex $(0,1)$, the director ${\hat{n}}$ winds by
$2\pi$. (c) In a disclination+half-vortex $(\half,\half)$, $\hat{n}$
and $\theta_c$ wind by $\pi$ simultaneously.
%, with no winding in
%$\varphi_+$. Thus, $\psi_{-}$ is suppressed at the vortex core,
%leading to a net spin moment along $+\hat{z}$ (black arrow).
}\label{fig:defects}
\end{figure}

More generally, the
topological point defects of a planar polar condensate in $2d$ are
labeled by two half-integer charges, $(q_c,q_s)$, describing the
winding of $\theta$ and $\phi$, respectively, in units of $2\pi$.
By single-valuedness of ${\bf \Psi}$ in Eq.~(\ref{eq:planar}), the sum $q_c+q_s$ is
constrained to be an integer. The lowest energy defects
are the superfluid vortex $(\pm
1,0)$, the nematic vortex $(0,\pm 1)$, and the
disclination+half-vortex $(\pm\half,\pm \half)$.  These defects are shown in Fig.~\ref{fig:defects}.

Topological defects proliferate when the temperature is large relative to some appropriate stiffness
parameter.  In the case at hand, due to the presence of both spin and charge degrees of
freedom, there are two relevant stiffness coefficients, $K_s$ and $K_c$.  These correspond
to the energy cost of an elastic deformation in the nematic direction $\phi$ and in the superfluid
phase $\theta$, respectively. In a dilute gas and in the absence of
an optical lattice, the kinetic term $\frac{\hbar^2}{2m} |\nabla
\Psi_\sigma|^2=$$\frac{\hbar^2}{2m} |\Psi|^2
\{(\nabla\theta)^2+(\nabla\phi)^2\}$ predicts
$K_c=K_s$ \cite{MukerjeeXuMoore}.  On the other hand, enhanced
quantum fluctuations in an optical lattice can change this
\cite{Barrett}.  An optical lattice suppresses both $K_c$ and
$K_s$, but its main effect is to impede atomic motion, leading to
$K_c/K_s<1$. For strong lattice potentials at integer filling, {\em i.e.} in
the Mott nematic phase, the charge stiffness is suppressed to the
point where the system is an insulator, while maintaining nematic
QLRO \cite{Imambekov}.  The nematic Mott insulator-polar SF
transition is second order. Thus, proximity to this transition
allows tuning the ratio $K_c/K_s\le 1$ over a wide range.

The topological defects interact logarithmically at long distances, leading to a
Coulomb gas action,
\begin{eqnarray}
S=\sum_{ij} \left(K_c  q^c_i q^c_j+K_s q^s_i q^s_j\right)
\log\frac{r_{ij}}{a}+\sum_i \log y_i\label{eq:CG}
\end{eqnarray}
Here, $a$ is the defect core size, and the reduced stiffnesses $K_c=\frac{\pi \rho_c}{2T}$ and $K_s=\frac{\pi \rho_s}{2T}$ have been normalized by a factor of
$2/\pi$ for later convenience.  The defect fugacity $y_i$ takes
the values $y_c$ and $y_m$, respectively, for the defects $(\pm 1,0)$ and $(\pm \frac12,\, \pm \frac12)$
in Fig.~\ref{fig:defects}.

\section{Defect unbinding}
\label{sec:instability}

The phase diagram as a function of the reduced stiffnesses $K_s$ and $K_c$ consists of four phases:  {\em (i)  Polar state.--}  At large $K_s$ and $K_c$, all defects remain bound and the polar order parameter has algebraic order.  {\em (ii)  Disordered.--} In the opposite limit, when both $K_s$ and $K_c$ are small, all of the topological defects proliferate, and the system has short-range correlations in both charge and spin.  {\em (iii) Nematic.--} Starting in the polar state and keeping $K_s$ large, when $K_c$ is reduced sufficiently, superfluid vortices proliferate, with all other defects remaining bound.  This leads to algebraic order in the nematic order parameter $e^{2i\phi}$, but no superfluidity.  {\em (iv) Paired superfluid.--}  Conversely,  when $K_c$ is large and $K_s$ is small, the nematic vortex unbinds, leading to algebraic order in $e^{2i\theta}$, but no spin order.  Note that, for polar condensates in an optical lattice, for which $K_s>K_c$, the paired superfluid is not present, as shown in Fig.~\ref{fig:TvsU}.

\begin{figure}
\includegraphics[width=3.2 in]{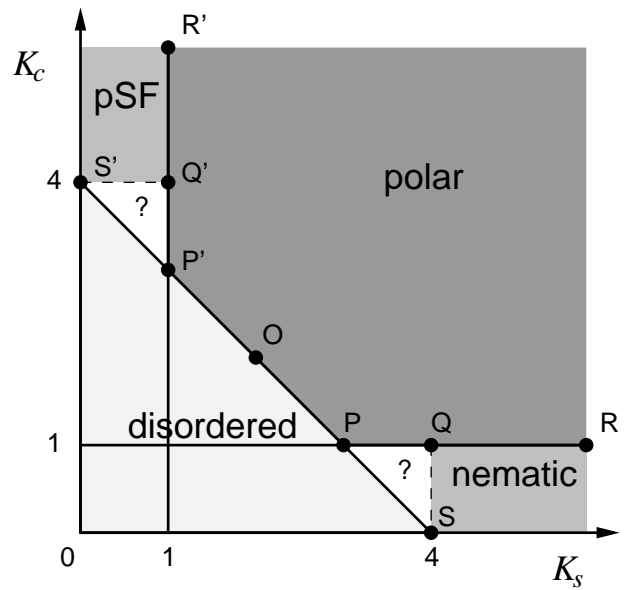}
\caption{Phase diagram as a function of the reduced stiffnesses $K_c$ and $K_s$.  The solid lines represent the conditions for the different topological defects to be bound, when the interactions between different types of defects are ignored.  Two regions (enclosed by {\bf PQS} and ${\bf P'Q'S'}$) are labeled by question marks.  In the absence of vortex interactions, these regions would be in the nematic and paired superfluid phases (pSF), respectively.  However, as argued in the text, the ordered phases in these two regions are unstable, and the two regions are part of the disordered phase, which extends to the dashed lines {\bf QS} and ${\bf Q'S'}$.   For simplicity, the phase diagram is drawn for infinitesimal bare defect fugacities.   Point {\bf O} is not a special point in the phase diagram, but it is indicated for comparison with Fig.~\ref{fig:TvsU}}\label{fig:conditions}
\end{figure}

The phase boundaries of the polar state can be obtained from the requirement that all defects in Fig.~\ref{fig:defects} be bound.  This corresponds to  $K_c^R>1$ (bound superfluid vortices),  $K_s^R>1$ (bound nematic vortices), and $K_c^R+K_s^R>4$ (bound disclination+half-vortices).  Here we have introduced the notation $K_{c,s}^R$ to denote the long-distance spin and charge stiffness, which is renormalized by the  presence of a finite density of bound defect pairs.  The conditions for vortex unbinding are shown in Fig.~\ref{fig:conditions}.

To obtain the correct phase diagram, however, it is important to recognize that {\em the conditions for defect unbinding cannot always be treated independently of one another}.  In particular, there are situations where the unbinding of one type of defect can precipitate the unbinding of a second type of defect.   As an example, consider the triangular region enclosed by {\bf PQS} in Fig.~\ref{fig:conditions}.  At first glance, it looks like the superfluid vortices are proliferated (since $K_c<1$) whereas the disclination+half-vortices remain bound (since $K_c+K_s>4$).  This would identify this region as belonging to the nematic phase.   
However, points within this region cannot describe a stable nematic phase.  The cheapest defects inside a nematic are the single disclinations -- in which the phase $\phi$ winds by $\pi$.  Disclinations are bound whenever the reduced spin stiffness $K_s$ is larger than 4.  However, $K_s<4$ in the region {\bf PQS}.  Therefore, this region is part of the disordered phase.

The instability of the nematic within region {\bf PQS} can also be understood by thinking of the single disclination as the remnant of the disclination+half-vortex of the polar phase, once the superfluid phase has been disordered due to the proliferation of superfluid vortices.  In this situation, the proliferation of superfluid vortices renormalizes the charge stiffness to zero ($K_c^R=0$).  Therefore, although the bare values of $K_c$ and $K_s$ satisfy the condition for disclination+half-vortex to be bound, $K_c+K_s>4,$ the renormalized stiffnesses do not, $K_c^R+K_s^R<4.$
These arguments show that, along the line {\bf PQ} (and also along ${\bf P'Q'}$), the system undergoes a direct transition from the polar state to the disordered state.  This transition involves two different types of defects, and therefore is in a different universality class from the transition along ${\bf P'OP}$.  We call the transition along {\bf PQ} a ``cascaded Kosterlitz-Thouless'' transition (cKT).  

It is difficult to give a fully controlled renormalization group (RG) treatment of the cKT transition, since the physics involves the proliferation of one type of defect before the other defect ``realizes'' that it is unbound.  The conventional RG treatment of defect unbinding is only controlled when the defect fugacities are small, and it breaks down when the defects proliferate and the fugacities become large.  Note that, despite the difficulty in giving a carefully controlled RG treatment, the coarse graining process in the RG cannot {\em increase} the value of the stiffnesses.  Therefore the argument for the instability of nematic order within the region {\bf PQS} is robust. In the following section, we will study the RG equations and derive the critical properties of the cKT transition {\bf PQ}, and we will contrast it to the disclination+half-vortex proliferation along ${\bf P'OP}$.  This will be followed in Sec.~\ref{sec:MC} by a numerical study of the transitions along {\bf PQ} and ${\bf P'OP}$ using Monte Carlo simulations.  This will allow us to study the finite size scaling properties at the two transitions, and compare them with the predictions from the RG analysis.

\section{Renormalization group analysis}
\label{sec:RG}

\subsection{Renormalization group equations}

In this section we focus on the case $K_s\ge
K_c$, corresponding to bosons in an optical lattice. Then, the
nematic vortices $(0,1)$ are always the last defects to proliferate, and they can
be neglected in the analysis below. A real space renormalization group
(RG) analysis of the Coulomb gas (\ref{eq:CG}) is carried out in
Ref.~[\onlinecite{KS}]. To quadratic order in the fugacities,
\cite{KS}%\cite{nienhuis}
\begin{subequations}\label{eq:RG}
\begin{align}
\dot{y}_c=&2(1-K_c)y_c+2\pi y_m^2\label{eq:ytRG}\\
%\dot{y}_s=&2(1-K_s)y_s+2\pi y_m^2\label{eq:yfRG}\\
\dot{y}_m=&%\left(4-K_s-K_c\right)y_m+2\pi
(4-K_s-K_c)y_m/2+2\pi y_m y_c\label{eq:ymRG}\\
%2\pi y_m(y_s+y_c)\label{eq:ymRG}
\dot{K}_c=&-8\pi K_c^2(2y_c^2+y_m^2)\label{eq:KtRG}\\
\dot{K}_s=&-8\pi K_s^2 y_m^2\label{eq:KfRG}
%\dot{K}_s=&-8\pi K_s^2(2y_s^2+y_m^2)\label{eq:KfRG}
\end{align}
\end{subequations}
Here $\dot{g}=\frac{dg}{d\ell}$ and $e^\ell$ is the length
rescaling factor. These RG equations are valid provided the
fugacities remain small. As is clear from Eq.~(\ref{eq:CG}), the
RG equations must be symmetric under the exchange of spin and
charge degrees of freedom.

In order to understand the RG equations (\ref{eq:RG}), consider first the flow of the superfluid vortex fugacity $y_c$,
Eq.~(\ref{eq:ytRG}). The first term on the RHS describes the
competition between energy cost and entropy gain for the creation
of a superfluid vortex; the second term, proportional to $y_m^2$,
arises because two dislocation+half-vortices can combine into a
superfluid vortex. On the other hand, Eq.~(\ref{eq:KtRG}) for the
flow of $K_c$ describes the screening of the Coulomb interaction
between defects due to a finite density of bound
defect-anti-defect pairs in the medium. The sign of the flow of
$K_c$ is negative semi-definite, and is only zero when the
fugacities $y_c$ and $y_m$ are zero, or when $K_c$ itself is zero.
Thus, the fixed point in the RG requires either the fugacities or
$K_c$ to flow to zero.  Similar considerations apply to $K_s$. The
fixed points of the RG are characterized by the values of the
renormalized stiffness $K_\gamma^R\equiv K_\gamma(\ell=\infty)$,
($\gamma=c,s)$.

The continuous transitions between the phases in
Fig.~\ref{fig:TvsU} arise from defect unbinding, and  may
be classified according to the type of defect that triggers the
transition. If a single type of defect is important, %at the transition,
one observes a conventional KT scenario. However, we also find a
class of transitions where unbinding of one set of defects triggers
the instability in another set, leading to a cascaded KT transition
with two diverging length scales.  Here, we will concentrate on the
two direct transitions between the polar and disordered states,
along the lines ${\bf OP}$ and {\bf PQ}.  The multicritical point
{\bf P} belongs to a different universality class, as discussed in
Ref.~[\onlinecite{KS}]. The other transitions {\bf QR}  and {\bf QS} belong
in the conventional KT scenario. % \cite{XYAT}.

\subsection{Disclination+half-vortex unbinding (${\bf OP}$)}   

Here, the
polar state is disordered by the proliferation of
disclination+half-vortices. These defects destroy both charge and
spin order. At the transition, the renormalized stiffness satisfy
$K_s^R+K_c^R=4$ and $K_c^R>1$. The superfluid
stiffness jump in this transition can be tuned {\em continuously}
with optical lattice depth, ranging from $K_c^R=2$ at point {\bf O}, to $K_c^R\rightarrow 1$ as we approach 
point {\bf P}.  On the other hand, the sum of
superfluid and spin stiffness is universal. The correlation
lengths $\xi_\gamma$ diverge as the transition is approached from
the disordered side as
\begin{eqnarray}
\xi_\gamma \sim a
\exp\left[d_\gamma(T-T_{KT})^{-\frac12}\right].\label{eq:corr}
\end{eqnarray}
As in the usual KT transition, $d_c=d_s$ is non-universal.

\subsection{Cascaded KT criticality ({\bf PQ})}

Along  {\bf PQ},
the superfluid vortices are on the verge of proliferating, since
$K_c^R=1$.  On the other hand, the sum $K_s^R+K_c^R$ is above the
threshold value of 4, indicating that the disclination+half-vortices
are bound at the transition. However, as soon as the transition is
crossed, $K_c$ flows to zero, reducing $K_s^R+K_c^R$ below $4$. Now,
the disclination+half-vortices unbind, leading to a completely
disordered phase.  We call this a ``cascaded'' KT (cKT) transition,
since unbinding of one type of defect triggers the unbinding of
the other.

Both spin-nematic and charge orders have diverging correlation
lengths as {\bf PQ} is approached from the disordered phase.
However, there is a separation of scales $\xi_s \gg \xi_c$, due to
the fact that the superfluid vortices unbind at a shorter length scale
than the disclination+half-vortices. We will show below that the
two are related by a power law,
\begin{eqnarray}
\xi_s \sim a (\xi_c/a)^B \label{eq:xiRat},
\end{eqnarray}
where $B=1/(4-K_s^R)>1$, and $\xi_c$ follows Eq.~(\ref{eq:corr}).

The main challenge in studying the cascaded KT transition is that
the naive RG equations (\ref{eq:RG}) break down once the
superfluid vortices unbind. In order to circumvent this, we use
the separation of scales to perform the RG in two steps.  First,
we solve Eq.~(\ref{eq:RG}) up to the scale $\xi_c$ where the
superfluid vortex fugacity begins to diverge.  At this point, the
superfluid correlations are explicitly short-ranged, and we can
integrate out the charge degrees of freedom to obtain a local
spin-only model. We then study the RG flow of the ensuing spin
model.

In the first step, the fugacity of the disclination+half-vortex is
renormalized down according to Eq.~(\ref{eq:ymRG}), $\tilde{y}_m\sim
y_m (a/\xi_c)^{(K_s-3)/2}$. The RG flow of the coarse grained
couplings $\tilde{y}_m$, and $\tilde{K}_c$ at longer scales is then
governed by $\dot{\tilde{y}}_m = 
\half\left(4-\tilde{K}_s\right)\tilde{y}_m$
and $\dot{\tilde{K}}_s =-8\pi \tilde{K}_s^2 \tilde{y}_m^2$.
Integrating this RG flow until $\tilde{y}_m$ is of order unity
yields Eq.~(\ref{eq:xiRat}), with $B=(4-K_s^R)^{-1}$. Note that $B$
is bounded below by one, near point {\bf P}, and can be arbitrarily
large near {\bf Q}, where $K_s^R=4$. The topology of the phase
diagram in Fig.~\ref{fig:TvsU} is crucially different from
Ref.~[\onlinecite{KS}], where the cKT transition ({\bf PQ}) is
misinterpreted as two separate transitions.

The {\em two} diverging length scales at the cKT transition arise
due to the fact that the disclination+half-vortex is a dangerously
irrelevant operator at the transition. This is one of the few
examples we know of a dangerously irrelevant {\em disorder}
operator \cite{OstlundHalperin,Senthil_science}.

\begin{figure}
\includegraphics[width=3.4 in]{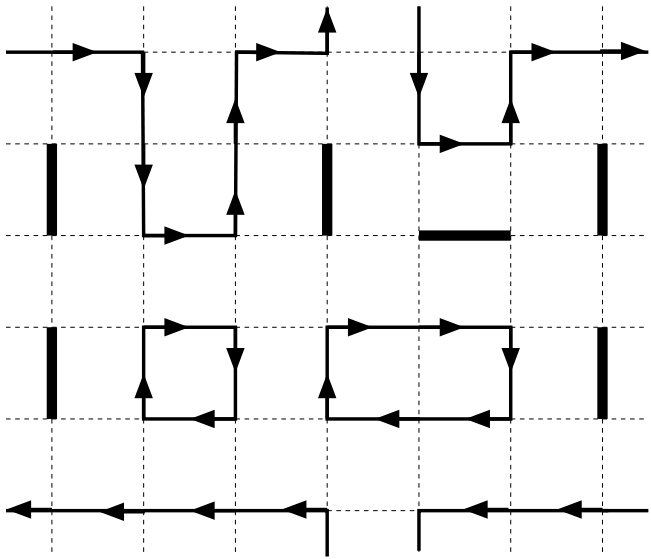}
\caption{An illustration of a current loop configuration discussed
in the text. The bonds with arrows are the directed dimers
and solid undirected bonds are the undirected double dimers. The 
configuration shown is planar, while in the model studied here 
the configurations will be similar but on a cubical lattice.\label{conf}}
\end{figure}

\section{Monte Carlo simulations}
\label{sec:MC}

\subsection{Loop model}

In order to test the new transitions predicted in 
the previous sections, we introduce a relatively simple 
lattice current-loop model defined on a periodic cubic 
lattice.  The configurations 
$[{\cal C}]$ that contribute to its partition function 
contain two types of bond variables: directed dimers 
$e_{i,\alpha}\in\{1,0,-1\}$ and undirected double 
dimers $d_{i,\alpha}\in\{0,2\}$, on bonds between site $i$ 
and its neighbors $i+\hat{\alpha}$ ($\alpha = x,y,\tau$). At 
each site $i$ we enforce two constraints: (1) 
$\sum_\alpha \{e_{i,\alpha}+e_{i-\hat{\alpha},\alpha}\}=0$ 
(directed dimer conservation) and (2) $\sum_\alpha
\left\{|e_{i,\alpha}|+d_{i,\alpha}+|e_{i-\hat{\alpha},\alpha}|
+d_{i-\hat{\alpha},\alpha}\right\}=2$ (close packing constraint). 
The partition function is given by
\begin{equation}
Z = \sum_{\cal C} \prod_i
\left\{W_D\right\}^{(d_{i,x}+d_{i,y}+d_{i,\tau})/2}
\left\{t\right\}^{|e_{i,\tau}|+d_{i,\tau}}.
\label{eq:loops}
\end{equation}
Roughly, the parameter $t$ is a temperature-like parameter 
(raising $t$ eliminates in-plane loops and double dimers), 
while $W_D$ tunes the ratio $K_c/K_s$ (increasing $W_D$ increases 
the double dimer density). As an illustration we show a planar 
lattice configuration in figure \ref{conf}. Configurations that 
contribute to the partition function are similar configurations 
but on a cubic lattice of size $L\times L\times 4$.

The loop model (\ref{eq:loops}) is a simple variant of strongly 
coupled two color 
lattice QCD with staggered fermions (SCLQCD2), a model with an 
$SO(3)\times U(1)$ symmetry and hence of interest also in the field 
of spinor condensates. In the model we consider here, the $SO(3)$
symmetry is broken to an $SO(2)$ subgroup. While SCLQCD2 has 
been studied in both cubic (3d) \cite{Chandrasekharan1} and 
hyper-cubic (4d) \cite{Chandrasekharan2} lattices, the above model 
remains unstudied so far. These models can be studied efficiently 
using directed-loop Monte Carlo algorithms as discussed 
in Refs.~\onlinecite{Chandrasekharan2} and\, \onlinecite{Chandrasekharan3}.

As a consequence of the constraints, the model considered here 
has two conserved currents: (1) $J_{i,\alpha}^s=\eta_i\{|e_{i,\alpha}|+d_{i,\alpha}-1/3\}$ where $\eta_i=+1$ and $-1$ on alternating sites,
and (2) $J_{i,\alpha}^c=e_{i,\alpha}$.
These currents correspond to the spin and charge conservation 
and can be used to compute $K_s$ and $K_c$:
\begin{equation}
K_\gamma=\frac{\pi}{2L^2}\left\langle \left(\sum_i J^a_{i,x} 
\right)^2\right\rangle.
\end{equation} 
Here, $K_\gamma$ is normalized such that at
a usual KT transition one would expect $K_s,K_c=1$.  Note that at
every space-like link, $J^c_{i,\alpha}+J^s_{i,\alpha}$ is an {\em even}
integer \footnote{Strictly speaking, $J^c_{i,\alpha}+J^s_{i,\alpha}$ is not an integer due to the staggered factor $\eta_i/3$.  However, the topological defects are determined from the sum of $J^c_{i,\alpha}+J^s_{i,\alpha}$ over the four values of the time-like coordinate at fixed space coordinates.  The staggered factor drops out from the sum, which is then an even integer.}.  This implies that the topological defects are half-integers with the constraint that $q_c+q_s$ is an integer, which is crucial to the physics here \footnote{Models based on U(1)$\times$U(1) continuous symmetry, but with different discrete symmetries than discussed here, have been shown to display different critical properties, as discussed {\em e.g.} in Ref.~\onlinecite{hasenbusch}}. In Appendix~\ref{app:Grassmann} we provide a diquark representation of the loop model, in which the symmetries of the model are seen explicitly. Due to these symmetries, the model studied here  is expected to exhibit the same universal physics as the Coulomb gas system described by Eq.~(\ref{eq:CG}) close to second order phase transitions. In fact we will provide below clear numerical evidence for the predicted transitions along the line {\bf OP} (disclination+half vortex unbinding) and {\bf PQ} (cKT transition) using this model.

\begin{figure*}[t]
\begin{center}
\hbox{
\includegraphics[width=0.47\textwidth]{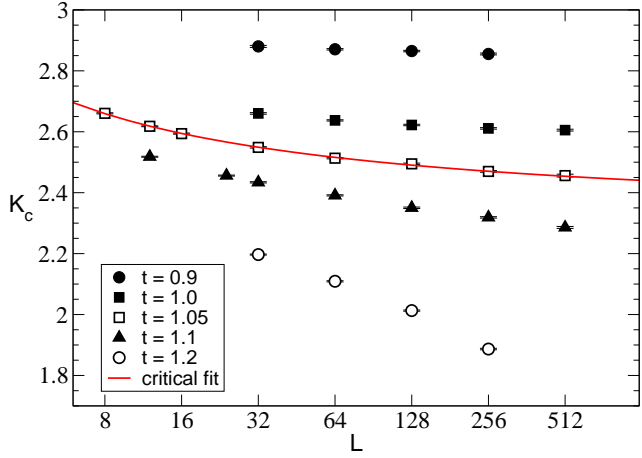}
\hskip0.3in
\includegraphics[width=0.47\textwidth]{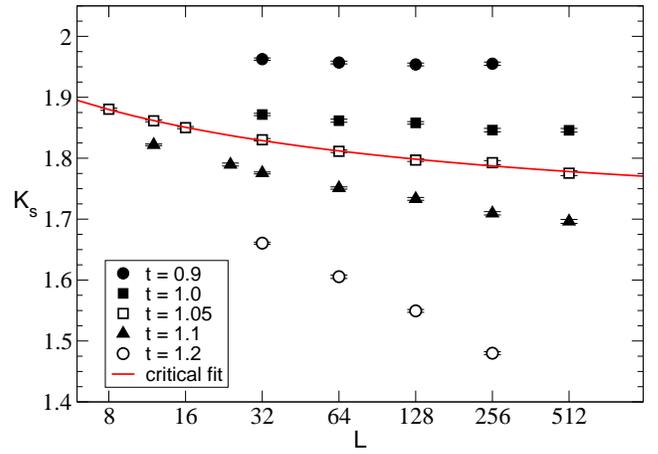}
}
\end{center}\
\caption{Finite size scaling near the polar melting transition
({\bf P'OP}) at $W_D=0$ and $t=0.9,1.0,1.05,1.1$ and $1.2$. The 
solid line is the fit to Eq.~(\ref{eq:finite4}). The values of 
the fit are given in table \ref{tab:PPprime}}\label{fig:PPprime}
\end{figure*}

\begin{table}[b]
\begin{center}
\begin{tabular}{|c||c|c|c||c|c|c||c|}
\hline  
$t$  & $K_c^R$&$C_c$&$\chi^2$&$K_s^R$&$C_s$&$\chi^2$&$K_s^R+K_c^R$ \\
\hline 
\multicolumn{8}{|c|}{$W_D=0$} \\
\hline
0.80&2.96(2)&7(1)&0.3&1.95(2)&$\infty$&0.7&4.9(4)\\
0.90&2.74(1)&6.5(9)&0.3&1.89(1)&10(2)&0.9&4.63(2)\\
1.00&2.468(5)&3.1(2)&1.6&1.763(5)&4.7(5)&1.1&4.231(10)\\
1.05&2.297(2)&1.09(3)&0.9&1.677(2)&2.05(7)&0.7&3.974(4)\\
1.10&2.145(2)&0.34(2)&24.7&1.594(2)&0.98(4)&4.4&3.739(4) \\ 
\hline
\multicolumn{8}{|c|}{$W_D=1$} \\
\hline
0.80&2.651(1)& $\infty$ &1.4&2.652(1)& $\infty$ &1.3&5.316(3)\\
0.90&2.41(2)&15(5)&0.6&2.40(1)&12(3)&1.1&4.81(3)\\
1.00&2.138(6)&4.2(4)&1.3&2.144(9)&4.6(7)&2.3&4.282(15)\\
1.05&1.967(2)&1.29(3)&2.6&1.968(2)&1.29(5)&2.3&3.935(4) \\
1.10&1.821(1)&0.43(2)&59&1.818(2)&0.40(2)&24&3.639(3) \\ \hline
\end{tabular}
\caption{Fits to Eq.~(\ref{eq:finite4}) at $W_D=0$ and $W_D=1$.
Note that $t=1.05$ is almost on the {\bf P'OP} line, with 
$K_s^R+K_c^R\approx 4$ with goodness of fit $\chi^2/DOF \approx 1-2$. When $W_D=1$
and $t=1.05$ we are at the midpoint {\bf O} on the {\bf P'OP} line.
\label{tab:PPprime}}
\end{center}
\end{table}

\subsection{Disclination+half-vortex proliferation (${\bf P'OP}$)}

Let us first focus on the transition that occurs along the line {\bf P'OP}.
When one is exactly on this line, the finite size scaling formula for $K_a$, 
can be computed using RG and is given by
\begin{eqnarray}
K_a (T_{KT},L)&=&K_a^R\left(1+\half\frac{1}{C_a+\log L}\right),
\label{eq:finite4}
\end{eqnarray}
where $K_c^R+K_s^R=4$ and the $C_a$ ($a=c,s$) are non-universal constants.
We would like to show evidence for this using our model. We have discovered 
that in our model we can approach this line by varying $t$ for fixed $W_D$ 
in the interval $0\le W_D<3$. We have done extensive calculations up to 
lattice sizes of $L=512$ by focusing on $W_D=0$ and $1$. In Fig.~\ref{fig:PPprime} 
we show the finite size scaling of $K_s$ and $K_c$ at $W_D=0$ for various values 
of $t$ close to the transition. Table~\ref{tab:PPprime} shows that we obtain a 
good fit to Eq.~(\ref{eq:finite4}) for $t \leq 1.05$. This is understandable 
since everywhere inside the polar phase $K_a$ is expected to be a constant 
for large $L$ and this just means $C_a$ is large. Since $t=1.05$ is the last 
value of $t$ where the fit works well, at that value we should be very close 
to the critical line. It is important to note that indeed $K_c+K_s \approx 4$ 
there (see last column of table \ref{tab:PPprime}). Individually each of the 
$K_a$'s are not universal. Since $K_c^R > K_s^R$ we must be somewhere on the 
{\bf P'O} line of the phase diagram. We can change $WD$ and $t$ to go to a 
different point on the {\bf P'OP} line. To show this have performed calculations 
at $WD=1$. In table \ref{tab:PPprime} we also give some of our fits for the 
$WD=1$ case. Again $t = 1.05$ seems to on the critical line and $K_c+K_s \approx 4$.
But now we have $K_c = K_s$ which means we are right at the midpoint {\bf O}.

\begin{figure*}[t]
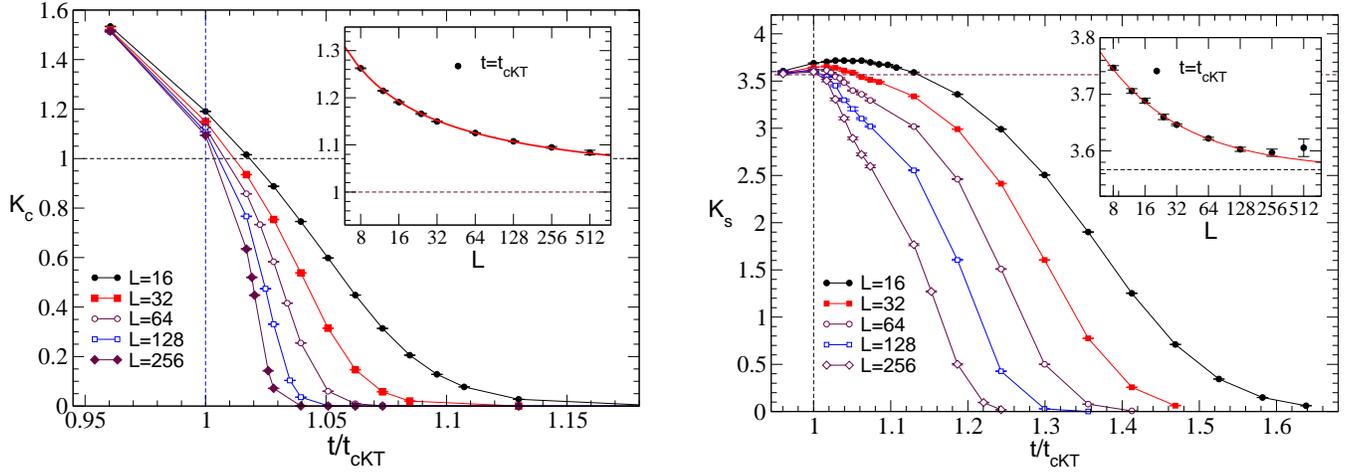

\begin{center}
\hbox{
\includegraphics[width=0.47\textwidth]{Kc}
\hskip0.3in
\includegraphics[width=0.47\textwidth]{Ks}
}
\end{center}
\caption{Cascaded KT ({\bf PQ}): $K_c$ (top) and $K_s$
(bottom) vs. $t/t_{cKT}$ for various system sizes $L$. {\bf
Insets:} $K_c$ and $K_s$ vs. $L$ at fixed $t=t_{cKT}$, fit to
Eqs.~(\ref{eq:finite3}) and (\ref{eq:finite2}),
respectively.  The fits yield $K_s^R=3.567(3),$   % with $\chi^2/DOF=1$
$C_s=0.58(2)$, and $C_c=-0.168(4)$, with % $\chi^2/DOF = 0.7$.
goodness of fit $\chi^2/DOF \sim 1$.  \label{fig:BFfiniteSize} 
}
\end{figure*}

\subsection{Cascaded KT transition ({\bf PQ})}

Next we focus on the cascaded KT transition line {\bf PQ}. On this line 
the cKT scenario predicts that $K_s$ and $K_c$ should follow different 
finite size scaling laws:
\begin{eqnarray}
K_c (t_{cKT},L)&=&1+\half\frac{1}{C_c+\log L},\label{eq:finite3}\\
K_s (t_{cKT},L)&=&K_s^R+C_s L^{3-K_s^R},\label{eq:finite2}
\end{eqnarray}
where $C_\gamma$ are again non-universal constants, and $3<K_s^R<4$.
In order to test this scenario, the first step is to locate a
point on the cKT line within our model. We
have identified numerically that one such point is $W_D=3.05$ and 
$t=t_{cKT}\equiv 0.885$. In order to cross the transition line at this
point we vary both $W_D$ and $t$ such that $W_D = 3.05 + 10(t-0.885)$, 
and compute $K_s$ and $K_c$ for lattice sizes up to $L=512$. 

Figure~\ref{fig:BFfiniteSize} shows our data for both $K_c$ and $K_s$.
The first important qualitative observation we make from the figure 
is that both $K_s$ and $K_c$ appear to jump to zero for $t > t_{cKT}$ 
as expected. Further the rough value of the jump is
$K_c^R \approx 1$ and $K_s^R\approx 3.6$ which is again as expected.
Thus, the point is roughly midway between {\bf P} and {\bf Q}, yet 
in the numerical data, both $K_s$ and $K_c$ seem to undergo a 
transition {\em simultaneously}.  This is consistent with 
the results of Sec.~\ref{sec:instability}, where we argued that for 
these parameters a split transition {\em i.e.}~crossing {\bf QR} first 
and then {\bf QS} is ruled out since $K_s^R<4$ is inconsistent with 
a stable nematic phase.

Next we perform a more quantitative analysis at $t=t_{cKT}$. We fit our
data for $K_c$ and $K_s$ as a function of $L$ to Eqs.~(\ref{eq:finite3})
and (\ref{eq:finite2}). If $K_c^R=1$ is fixed we 
find $C_c=-0.168(4)$, with goodness of fit $\chi^2/DOF=1$, and $K_s^R=3.567(3),$ 
and $C_s=0.58(2)$ with $\chi^2/DOF = 0.7$. We emphasize that 
$ 8 \leq L \leq 512$ were used in the fit. This large range gives 
us confidence in our analysis. The insets in the graphs in 
Fig.~\ref{fig:BFfiniteSize} show the finite size scaling 
of $K_c$ and $K_s$, respectively along with the fit. Thus, we
claim that the Monte Carlo simulations are consistent with the 
expected finite size scaling predicted by the RG analysis.

Another important prediction of the cKT scenario is the presence of
two diverging correlation lengths near the cKT transition {\bf PQ}, 
which follows Eq.~(\ref{eq:xiRat}). Unfortunately we have not measured
the correlation lengths directly. However we can see this divergence
in an indirect manner. Since a finite size box limits 
the diverging correlation lengths, the jumps in the stiffness $K_a$ 
are no longer sharp but broadened depending on the value of $L$ 
(see Fig.~\ref{fig:BFfiniteSize}). A temperature $t^*_a > t_{KT}$ 
can then be defined such that $K_a (t^*_a)=x K_a^R(t_{KT})$ where 
$x < 1$ is some fixed fraction. In other words, at $t^*_a$, $K_a$ 
has been reduced to a fraction $x$ of its value at $t_{KT}$. The 
fact that the correlation length diverges as $t$ approaches 
$t_{KT}$ is now seen by the fact that as $L$ becomes large $t^*_a$ 
approaches $t_{KT}$. This behavior can be quantified using RG and 
we expect
\begin{eqnarray}
\frac{1}{\sqrt{t^*_a-t_{KT}}}=\alpha_a \log L+\beta_a
%+\frac{\gamma_a}{(\alpha_a\log L+\beta_a)^2}
\label{eq:finite1} 
\end{eqnarray}
where all constants are non-universal, except for the ratio
$\alpha_c/\alpha_s=1/(4-K_s^R)$. Note that Eq.~(\ref{eq:finite1}) 
is valid provided that $t_a^*-t_{KT}$ is small, {\em i.e.} for 
large enough system size $L$. 

\begin{figure}
\includegraphics[width=0.47\textwidth]{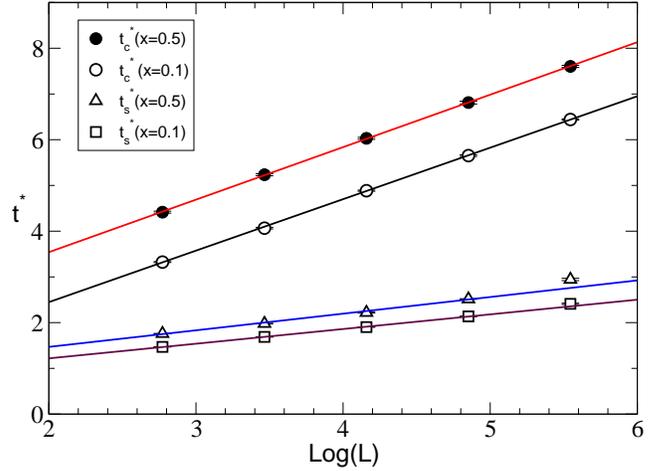}
\caption{$(t^*-t_{KT})^{-1/2}$ vs. system size for charge
correlations and spin correlations at $x=0.1$ and $x=0.5$.
The solid lines are fits given in table \ref{tab:tstar} \label{fig:tstar}}
\end{figure}

Fig.~\ref{fig:tstar} shows our data for $t^*_a$ for the choices $x=0.1$ and $x=0.5$.
The first striking qualitative observation we make is that $1/\sqrt{t^*_a-t_{KT}}$
scales linearly with $\log(L)$ to a very good approximation as expected. Further 
the slopes of the two lines are clearly different. Note that we expect 
$\xi_s \sim \xi_c^p$ where $p = {\alpha_c/\alpha_s} > 1$. Indeed we do find 
that $p > 1$. Quantitatively, while $t^*_c$ fits well to Eq.~(\ref{eq:finite1}),  
$t^*_s$ does not fit for larger values of $L$ due to a small but clear curvature 
in the data. Table \ref{tab:tstar} contains the fit results.
\begin{table}[h] 
\begin{center}
\begin{tabular}{|c|c|c|c||c|c|c|}
\hline  
$x$  & $\alpha_s$&$\beta_s$&$\chi^2/DOF$&$\alpha_c$&$\beta_c$&$\chi^2/DOF$ \\
\hline 
0.1&0.321(2)&0.577(8)&8.3&1.125(5)&0.20(2)&1.4 \\
0.5&0.363(3)&0.74(1)&35&1.15(1)&1.25(4)&0.3 \\
\hline
\end{tabular}
\caption{Fits to Eq.~(\ref{eq:finite1}) at $x=0.1$ and $x=0.5$. The fits for
$t_s^*$ are not very good because of a detectable curvature in the data.
\label{tab:tstar}}
\end{center}
\end{table}
Ignoring the large values of $\chi^2/DOF$, we see that $\alpha_c/\alpha_s \sim 3.5$
at $x=0.1$ and $\sim 3.2$ at $x=0.5$. These must be compared to the theoretical 
prediction from the RG treatment, $1/(4-K_s^R)=2.31(1)$. If we take the curvature
in $t^*_s$ into account by using a fit of the form 
$\frac{1}{\sqrt{t^*_s-t_{KT}}}=\alpha_s \log L+\beta_s + \gamma_s/(\log L)$ 
the $\chi^2/DOF$ improves slightly but is still not good. The ratio 
$\alpha_c/\alpha_s$ changes to about $3$ and $2.3$ at $x=0.1$ and $0.5$. 
If instead we just use the last two points and draw a straight line through the
data, this ratio is $2.8$ and $1.85$ respectively. This large variation 
in the values of the ratio $\alpha_c/\alpha_s$ shows that we do not yet have 
quantitative control on it. The theoretical expectation seems to be 
within the large systematic errors. Since our data are obtained on rather 
small lattice sizes where the two large length scales $\xi_s\sim\xi_c^{2.3}$ 
cannot fit well, our inability to quantitatively control the ratio is 
not surprising.

\section{Conclusion}
\label{sec:conclusion}

%{\bf Can be removed if we include in intro: The same is true for
%uniaxial nematic condensates of {\em odd} spin. On the other hand,
%for an {\em even} spin nematic condensate, disclinations do not
%induce a minus sign in the order parameter, and therefore the charge
%and spin degrees of freedom are not topologically intertwined.
%Furthermore, for a large enough system, an optical lattice may not
%be necessary to observe this effect: for weak quadratic Zeeman
%fields, thermal fluctuations of spin are stronger than those of
%charge, leading to $K_c/K_s>1$. However, unless the quadratic
%Zeeman field is extremely small, these fluctuations only create a
%weak anisotropy between charge and spin stiffness.
%}

%{\bf can be removed - no new info: We have studied the interplay of
%charge and spin degrees of freedom in polar spinor condensates which
%leads to a varied set of finite temperature critical phenomena in
%2D. }

In conclusion, we have shown that the topological binding of spin and charge 
vorticity in $S=1$ polar condensates can give new types of phase transitions.
In particular we have shown the existence of the disclination+half vortex unbinding
transition, which is similar to the KT transition but where the superfluid
stiffness jump is non-universal and a cascaded KT transition where in addition
to a non-universal jump in the stiffness two correlation lengths diverge at
the critical point, where one is a power of the other. A large scale numerical 
study supports the detailed picture we have developed. 

Our analysis can be applied to other ordered states. For example, 
spin-2 $^{87}$Rb in a magnetic field forms a ``square nematic''
state \cite{sengstock,turner} with $\Psi \sim 
e^{i\theta}(e^{2i\phi_s},0,0,0,e^{-2i\phi_s})$. This yields
precisely the same physics as the spin-1 planar polar state. 
There are examples outside of cold atomic systems which display the the same topological structure studied here.  For instance, Berg {\em et al.} have argued that the thermal melting of a striped superconducting state is produced by a fully analogous set of topological defects to those discussed here\cite{berg}.  Therefore, the phase diagram of a striped superconductor is the same as that of a planar polar condensate, although the particular phases involved in the two cases are different.  Following this work, a recent numerical study has looked at the $3d$ version of this system \cite{BabaevSudbo}.

We thank E.~Berg, G.~Delfino, J.E.~Moore, S.~Mukerjee, L.~Radzihovsky, and
D.~Stamper-Kurn for useful discussions.  This work was supported in
part by the NSF grant DMR-0506953 and the Hellman Faculty Fund.

\appendix

\section{Diquark representation of the loop model}
\label{app:Grassmann}

In this Appendix, we recast the loop model Eq.~(\ref{eq:loops}) in terms of a path integral over Grassmann variables in order to clarify the symmetries of the model. The partition function is given by
\begin{eqnarray}
Z_{\rm G} = \int {\cal D}\psi_\uparrow{\cal D}\psi_\downarrow {\cal D}\overline{\psi}_\downarrow{\cal D}\overline{\psi}_\uparrow e^{-S_{\rm G}}
\label{eq:grass}
\end{eqnarray}
where the action
%\begin{eqnarray}
%S_{\rm G}&=&\sum_{i,\alpha}u_\alpha \left(\overline{\psi}_{i\uparrow}\overline{\psi}_{i\downarrow}\psi_{i+\hat{\alpha}\downarrow}\psi_{i+\hat{\alpha}\uparrow}+\overline{\psi}_{i+\hat{\alpha}\uparrow}\overline{\psi}_{i+\hat{\alpha}\downarrow}\psi_{i\downarrow}\psi_{i\uparrow}\right)\nonumber\\
%&-&(W_D-1)\sum_{i,\alpha} u_\alpha^2 \overline{\psi}_{i\uparrow}\overline{\psi}_{i\downarrow}\psi_{i\downarrow}\psi_{i\uparrow}\overline{\psi}_{i+\hat{\alpha}\uparrow}\overline{\psi}_{i+\hat{\alpha}\downarrow}\psi_{i+\hat{\alpha}\downarrow}\psi_{i+\hat{\alpha}\uparrow}. \nonumber
%\end{eqnarray}
%\begin{eqnarray}
%Z_{\rm G} = \int {\cal D}\overline{\psi}_\uparrow {\cal D} \psi_\uparrow
%{\cal D}\overline{\psi}_\downarrow{\cal D}\psi_\downarrow  e^{-S_{\rm G}}
%\label{eq:grass}
%\end{eqnarray}
%where the action
\begin{eqnarray}
S_{\rm G}&=& - \sum_{i,\alpha} \Bigg[
\frac{u_\alpha}{2}\Big\{(\overline{\Psi}_{i+\alpha}\Psi_{i})^2 + 
(\overline{\Psi}_{i}\Psi_{i+\hat{\alpha}})^2\Big\}\
\nonumber\\
&&\ \ \ +\ (W_D-1) \left(\frac{u_\alpha}{2}\right)^2 
(\overline{\Psi}_{i+\alpha}\Psi_{i})^2
(\overline{\Psi}_{i}\Psi_{i+\hat{\alpha}})^2\Bigg]. \nonumber
\end{eqnarray} 
Here
\begin{equation}
\Psi_i \equiv \left(\begin{array}{c} \psi_{i,\uparrow} \cr \psi_{i,\downarrow}\end{array}\right),\ \ 
\mbox{and}\ \ 
\overline{\Psi}_i \equiv \left(\begin{array}{cc} 
\overline{\psi}_{i\uparrow} & \overline{\psi}_{i\downarrow}\end{array}\right)\ \ 
\end{equation}
such that 
$\psi_{i\sigma}$ and $\overline{\psi}_{i\sigma}$ ($\sigma=\uparrow,\downarrow$) are independent Grassmann variables residing on the sites of a $L\times L\times 4$ lattice. In particular, $\psi_{i\sigma}$ and $\overline{\psi}_{i\sigma}$ are {\em not} complex conjugates of each other. The constants $u_\alpha$ are $u_x=u_y=1$ and $u_\tau=t$, so that $\hat{\tau}$ can be thought of as an Euclidean time direction and $t$ as a temperature-like parameter.  
Performing the path integral over $\psi_{i\sigma}$ and $\overline{\psi}_{i\sigma}$ yields the loop model (\ref{eq:loops}) exactly, as can be checked easily by explicit computation.

Equation~(\ref{eq:grass}) can also be expressed as a model of ``diquarks'' $D_i\equiv\psi_{i\downarrow}\psi_{i\uparrow}$ and $\overline{D}_i\equiv\overline{\psi}_{i\uparrow}\overline{\psi}_{i\downarrow}$  hopping and interacting on a lattice,
\begin{eqnarray}
S_{\rm G}&=&\sum_{i,\alpha}u_\alpha (\overline{D}_i D_{i+\alpha}+\overline{D}_{i+\alpha}D_i)\nonumber\\
&-&(W_D-1)\sum_{i,\alpha} u_\alpha^2 \overline{D}_i D_i \overline{D}_{i+\alpha}D_{i+\alpha}. 	
\end{eqnarray}
The model has an SU(2)$\times$SU(2) gauge symmetry.  To see this, note that the diquark $D_i=\psi_{i\downarrow}\psi_{i\uparrow}=\frac{1}{2}\epsilon_{\sigma\sigma'}\psi_{i\sigma}\psi_{i\sigma'}$ is invariant under a {\em local} transformation
\begin{eqnarray}
\Psi_i  \ \to\  U_i \ \Psi_i,\label{eq:gauge}
\end{eqnarray}
where $U_i\in\mathrm{SU(2)}$.  Similarly, an independent SU(2) transformation $\overline{U}_i$ can be carried out on the barred variables, $\overline{\Psi}_i\to \overline{\Psi}_i\ \overline{U}_{i}^\dagger$, leaving $\overline{D}_i$ invariant.
%Equation (\ref{eq:grass}) is invariant under {\em local} transformations,
%\begin{equation}
%\Psi_i  \ \to\  U \ \Psi_i,\ \ \ \ 
%\overline{\Psi}_i\ \to\  \overline{\Psi}_i \ U^\dagger \label{eq:gauge}
%\end{equation}
%for $U\in SU(2)$.  In other words, the diquark model has a lattice SU(2) gauge invariance, which can be used to fix three of the relative phases between the four Grassmann variables $\psi_{\sigma}$ and $\overline{\psi}_\sigma$.  

After taking the gauge invariance into account, we are still left with an independent {\em global}  U(1)$\times$U(1) symmetry of the model (\ref{eq:grass}), parametrized by the angles $\theta$ and $\phi$,
\begin{eqnarray}
\Psi_{i}&\to& e^{i(\eta_i\phi-\theta)/2}\ \Psi_{i}\nonumber\\
\overline{\Psi}_{i}&\to&  e^{i(\eta_i\phi+\theta)/2}\ \overline{\Psi}_{i}
\label{eq:symm}
\end{eqnarray}
where $\eta_i=+1$ and $-1$ on alternating sites.  Note that, unlike the gauge invariance in Eq.~(\ref{eq:gauge}), the transformation (\ref{eq:symm}) does not leave the diquark invariant, since $\psi_{i\downarrow}\psi_{i\uparrow}\to e^{i(\eta_i\phi-\theta)} \psi_{i\downarrow}\psi_{i\uparrow}$.  However, a simultaneous shift of $\theta$ and $\phi$ by the angle $\pi$ does leave the diquark invariant.  Hence, the model (\ref{eq:grass}) is explicitly seen to have topological defects that are labeled by half-integers with the constraint that $q_c+q_s$ is an integer.  Therefore, the low energy defects of the model are the superfluid vortex $(1,0)$, the nematic vortex $(0,1)$, and the nematic+half-vortex $(\half,\half)$. By equivalence of the diquark model to Eq.~(\ref{eq:loops}), we conclude that the loop model also has the same topological defects.

\end{document}